\documentclass[12pt, twocolumn, english]{iopart}
\expandafter\let\csname equation*\endcsname\relax
\expandafter\let\csname endequation*\endcsname\relax
\usepackage{amsmath}
\usepackage[T1]{fontenc}
\usepackage[english]{babel}
\usepackage{bm}
\usepackage{bbm}
\usepackage{iopams}
\usepackage{amssymb}
\usepackage{bbold}
\usepackage{graphicx}
\usepackage{appendix}

\usepackage{color}

\usepackage[%
  breaklinks=true,
  colorlinks=true,
  linkcolor=blue,
  urlcolor=blue,
  citecolor=blue
]{hyperref}
 \usepackage[numbers,sort&compress]{natbib}
 \usepackage{hypernat}
 
\begin{document}

\title{Dynamics of the relativistic electron spin in an electromagnetic field}

\author{Ritwik Mondal$^1$, Peter M. Oppeneer$^2$}
\address{$^1$Fachbereich Physik and Zukunftskolleg, Universit\"at Konstanz, DE-78457 Konstanz, Germany}
\address{$^2$Department of Physics and Astronomy, Uppsala University, Box 516, SE-75120 Uppsala, Sweden}

\ead{ritwik.mondal@uni-konstanz.de}

\begin{abstract}
A  relativistic spin operator cannot be uniquely defined within relativistic quantum mechanics. Previously, different proper relativistic spin operators have been proposed, such as spin operators of the Foldy-Wouthuysen and Pryce type, that both commute with the free-particle Dirac Hamiltonian and represent constants of motion. Here we consider the  dynamics of a relativistic electron spin in an external electromagnetic field. We use two different Hamiltonians to derive the corresponding spin dynamics. These two are: (a) the Dirac Hamiltonian in presence of an external field, (b) the semirelativistic expansion of the same. Considering the Foldy-Wouthuysen and Pryce spin operators we show that these lead to different spin dynamics in an external electromagnetic field, which offers possibilities to distinguish their action. We find that the dynamics of both spin operators involve spin-dependent and spin-independent  terms, however, the Foldy-Wouthuysen spin dynamics additionally accounts for the relativistic particle-antiparticle coupling. We conclude that the Pryce spin operator provides a suitable description of the relativistic spin dynamics in a weak-to-intermediate external field, whereas the Foldy-Wouthuysen spin operator is more suitable in the strong field regime.  
\end{abstract}

\section{Introduction}
Spin, in quantum mechanics, is an intrinsic property of an elemental particle e.g., of the electron. However, in contrast to nonrelativistic quantum mechanics, the definition of the spin operator is not unique in relativistic quantum mechanics \cite{Newton1949,Jordan1963, Lorente1974,Bauke2014PRA,Bliokh2017}.
In nonrelativistic quantum mechanics, the spin is expressed by the Pauli spin matrices as $\bm{\sigma}$ and the corresponding spin angular momentum by $\bm{S} = \frac{\bm{\sigma}}{2}$ (assuming units such that $\hbar= 1$). The latter definition is valid for the two component Schr\"odinger or Pauli Hamiltonian that relates directly the spin operator to the Pauli spin matrices.
However, in a relativistic formulation the spin angular momentum cannot be defined separately because the total angular momentum has to be conserved. Therefore, the definition of spin angular momentum depends on the definition of the orbital angular momentum. Generally, the orbital angular momentum is defined as $\bm{L}=\bm{r}\times \bm{p}$ such that the total angular momentum is calculated as $\bm{J} = \bm{L} +\bm{S}$ in a nonrelativistic framework. Even so, in relativistic quantum mechanics, the position operator is not uniquely defined and, consequently, the spin angular momentum does not have a unique definition \cite{Newton1949,Jordan1963,OConnell1997}. In fact, for both the orbital and spin angular momentum 
several definitions have been proposed \cite{OConnell1997}. 

While the definition of the relativistic spin operator might seem a semantic issue, its formulation does in fact matter when relativistic spin dynamics is considered.
Spin dynamics has previously been computed starting from the nonrelativistic spin operator, \textit{i.e.}\ the Pauli spin matrices $\bm{\sigma}$ \cite{hickey09,Mondal2016}. The resulting equation of motion is found to be composed of spin precession, spin relaxation and even spin nutation (inertial dynamics), terms that are consistent with the well-known Landau-Lifshitz-Gilbert (LLG) equation of spin dynamics. Even though the LLG equation  has been used for the spin dynamics at ultrashort timescales, its applicability at these timescales has been questioned \cite{Atxitia2016}. However, the relativistic spin dynamics has not yet been derived from a relativistic spin operator. 
With this objective, we derive  in this article the spin dynamics for relativistic spin operators, in particular, we treat the previously proposed proper relativistic spin operators  due to Foldy-Wouthuysen \cite{foldy50,Foldy1952} and Pryce \cite{Pryce1935,Pryce1948}. We consider three different Hamiltonians to derive the  relativistic spin dynamics: (1) the free-particle Dirac Hamiltonian, (2) the Dirac Hamiltonian in an electromagnetic (EM) environment, (3) the Foldy-Wouthuysen (FW) transformation of the Dirac Hamiltonian in an EM environment. The results show that the corresponding spin dynamics leads to the LLG equation of motion, however with additional contributions due to relativistic spin operator formulations.
Comparing the relativistic dynamics for an electron spin in an EM field, we draw the conclusion that the Pryce spin operator provides a suitable formulation of the relativistic electron spin dynamics in the weak to intermediate field regime, however, the FW spin operator is more applicable for describing spin in the relativistic strong field regime.

 In the following we first introduce the relativistic spin operators, especially the FW and Pryce spin operators. Thereafter, in Sec.\ 3 we formulate the three different Hamiltonians that will be used to evaluate the relativistic spin dynamics. Then, in Sec.\ 4 we derive the spin dynamics corresponding to the FW and Pryce spin operators and discuss the obtained results. Conclusions are drawn in Sec.\ 5.

\section{Relativistic spin operators}

The free-particle Dirac Hamiltonian reads \cite{Dirac1928,Dirac1928a,Dirac1930}
\begin{align}
\mathcal{H}_{\rm D}^0 & = c\,\bm{\alpha} \cdot \bm{p} + \beta m_0c^2\,,
\label{free-Dirac}
\end{align}
with the rest mass $m_0$, and $\bm{\alpha}$ and $\beta$ are the $4 \times 4$ Dirac matrices which obey the following relations \cite{strange98} 
\begin{align}
\alpha_i^2 = \mathbb{1}, \qquad \beta^2 = \mathbb{1}, \qquad \alpha_i\alpha_j + \alpha_j\alpha_i = 2\delta_{ij}, \qquad \alpha_i\beta +\beta\alpha_i = 0 \,,  
\end{align}
where $\mathbb{1}$ is the $4\times4$ identity matrix.  
 The corresponding spin operator is a four-component operator that 
describes both the particle spin up (down) and antiparticle spin up (down) states.
The Dirac spin operator has hence the definition $\bm{S}_{\rm D} = \frac{ \bm{\Sigma}}{2}$, with the components of the operator $\bm{\Sigma}$ ($=\mathbb{1}\otimes\bm{\sigma}$) as
\begin{align}
\Sigma_j & = -i \alpha_k\alpha_l\,.
\end{align}
In addition, the orbital angular momentum is given as $\bm{L}_{\rm D} = \bm{r}\times\bm{p} $, which has to be multiplied by a  2-units 
block diagonal matrix of $2\times 2$. The total angular momentum is then given by $\bm{J}_{\rm D} = \bm{L}_{\rm D} + \bm{S}_{\rm D}$ and it recovers the angular momentum in the nonrelativistic framework when taken in two-component form.

The spin does not couple to the orbital angular momentum in nonrelativistic quantum mechanics, in fact, the spin operator $\bm{S}$ is a constant of motion when  the Schr\"odinger Hamiltonian is considered. However, for a free-particle Dirac Hamiltonian, the calculation of the spin dynamics with the Dirac spin operator reveals
\begin{align}
    \frac{d\bm{S}_{\rm D}}{d t} & = -c\,\bm{\alpha}\times \bm{p}\,,
\end{align}
 meaning that the Dirac spin operator is not a constant of motion. As one expects, the corresponding dynamics contains the particle-antiparticle coupling strength, following the feature of the Dirac Hamiltonian. Moreover, the dynamics suggests that the Dirac matrices $\alpha_i$ are coupled to the orbital angular momentum via $\bm{p} = -i\bm{\nabla}$. Furthermore, it has been shown that the eigenvalues of the Dirac spin operator deviate from $\pm 1/2$ for 
 materials having higher atomic numbers \cite{Bauke2014NJP}. The latter is understandable because, for higher atomic numbers, the spin cannot be considered as an independent quantity, rather the spin is coupled to the orbital degrees of freedom due to larger spin-orbit coupling. Thus, the two major drawbacks of the Dirac spin operator is that (a) it does not commute with the free-particle Dirac Hamiltonian, (b) the eigenvalue does not correspond to $\pm 1/2$ for systems with higher atomic number, which implies that the Dirac spin operator cannot be considered as a proper relativistic spin operator.        
  
A proper relativistic spin operator should have the following properties \cite{Bauke2014NJP,Bauke2014PRA}:
\begin{enumerate}

\item It has to commute with the relativistic free-particle Dirac equation. This implies that the spin operator is a constant of motion for a Dirac free-particle.
\item It has to obey the SU(2) algebra of spin operators. The commutator of two spin operators should follow the relation
\begin{align}
\left[ S_i,S_j\right] & = i \epsilon_{ijk} S_k\,, 
\end{align}
where $\epsilon_{ijk}$ is the anti-symmetric Levi-Civita tensor.
\item The spin operator must have two eigenvalues of $\pm \frac{1}{2}$.
\item The total angular momentum has to be conserved.
\end{enumerate}

There have been a number of relativistic spin operators reported in the literature \cite{Bauke2014NJP,Caban2013,Bauke2014PRA,Bliokh2017}. However, all of these existing spin operators do not obey all the aforementioned conditions.  
It is known that  {\it only} the relativistic Foldy-Wouthuysen and Pryce spin operators \cite{foldy50,Foldy1952,Pryce1935,Pryce1948} satisfy all the above-mentioned properties. Therefore, one can infer that they can be considered as proper spin operators \cite{Bauke2014NJP}. 

In the following, we calculate the spin dynamics corresponding to both of these operators for a system excited by an electromagnetic field (e.g., {a} laser pulse). 

\subsection{FW spin operator}
The FW spin operator has the following definition \cite{foldy50,DEVRIES1968,Foldy1952,strange98,greiner00}
\begin{align}
\bm{S}_{\rm FW} & = \frac{1}{2}\bm{\Sigma} + \frac{i\beta\,\bm{p}\times\bm{\alpha}}{2E_p}  - \frac{\bm{p}\times \left( \bm{\Sigma} \times \bm{p}\right)}{2E_p (E_p + m_0c^2)}\,,  
\label{FW_spin_operator}
\end{align}
with the energy $E_p = \sqrt[]{p^2c^2+m_0^2c^4}$. Correspondingly, the position operator is also defined as 
\begin{align}
\bm{r}_{\rm FW} & = \bm{r} - \frac{i\beta\bm{\alpha}}{2E_p} + \frac{i\beta\left( \bm{\alpha}\cdot\bm{p}\right)\bm{p} - \left(\bm{\Sigma}\times \bm{p}\right) \vert \bm{p}\vert }{2E_p(E_p+m_0c^2) \vert \bm{p} \vert}\,,
\end{align}
such that the total angular momentum is exactly the same as that of the nonrelativistic case i.e., $\bm{J}_{\rm FW} = \bm{L}_{\rm FW} + \bm{S}_{\rm FW} = {\bm{r}_{\rm FW} \times \bm{p} +\bm{S}_{\rm FW}} = \bm{r} \times \bm{p} + \frac{\bm{\Sigma}}{2}$. This {construction `made by hand'} reflects that the total angular momentum has to be conserved, and has to be equal to the total angular momentum for the Pauli representation when we consider the two-component form.   

\subsection{Pryce spin operator}
The Pryce spin operator has the following definition \cite{Pryce1935,Pryce1948}:
\begin{align}
    \bm{S}_{\rm Py} & = \frac{1}{2}\beta \bm{\Sigma} + \frac{1}{2} \left(1-\beta\right) \frac{(\bm{\Sigma} \cdot \bm{p})\bm{p}}{p^2}\, ,
    \label{Pryce_spin_operator}
\end{align}
and the corresponding position operator has the form
\begin{align}
    \bm{r}_{\rm Py} & = \bm{r} - \frac{1}{2}\left(1-\beta\right) \frac{\bm{\Sigma}\times \bm{p}}{p^2}\,,
\end{align}
such that the total angular momentum is written as $\bm{J}_{\rm Py} =\bm{L}_{\rm Py} + \bm{S}_{\rm Py} =  \bm{r}_{\rm Py} \times \bm{p} + \bm{S}_{\rm Py}  = \bm{r} \times \bm{p} + \frac{\bm{\Sigma}}{2}$. The derived total angular momentum for {the} Pryce spin and orbital momentum operator is equal to the total angular momentum in {the} Pauli representation as argued earlier. 

A striking difference between FW and Pryce spin operators is that FW spin operator contains a coupling term, i.e., the second term of Eq.\ (\ref{FW_spin_operator}), however, such coupling terms do not appear in the Pryce spin operator in Eq.\ (\ref{Pryce_spin_operator}). 
One immediately notices that the spin operators contain {\it not only} the spin angular momentum, {\it but also}, the orbital angular momentum in the form of $\bm{p}$. The same is valid for the position operators as well, because of the following reasons. For the FW and Pryce position operators, we obtain (neglecting higher-order terms)
\begin{align}
    r_{\rm FW}^2 & =  r^2 + \frac{1}{4E_p^2} + \frac{i\beta (\bm{r}\cdot \bm{p})(\bm{\alpha}\cdot \bm{p})}{E_p(E_p+m_0c^2)\vert \bm{p} \vert} + \frac{\bm{\Sigma}\cdot \bm{L}}{E_p(E_p+m_0c^2)}\,,\\
    r_{\rm Py}^2 & =  r^2 + \left(1-\beta\right)\frac{\bm{\Sigma}\cdot \bm{L}}{p^2}\,,
\end{align}
respectively. Here, the last correction terms represent the well-known spin-orbit coupling that is missing in a nonrelativistic description. Note that there is another relativistic correction term that appears in the FW position operator which is notably off-diagonal in the particle-antiparticle Hilbert space.
Having these proper relativistic spin operators, we derive their spin dynamics, particularly, in an applied EM field. While both operators are proper spin operators, their formulation is evidently different, and it is unknown which spin operator provides a more suitable description of the dynamics. In particular, we are keen to understand the effects of relativistic coupling terms within the corresponding spin dynamics.

\section{Relativistic Hamiltonians}
For deriving the spin dynamics, we consider three different Hamiltonians. The first one is the Dirac free-particle Hamiltonian that has already been introduced in Eq.\ (\ref{free-Dirac}). The second one is the Dirac equation in the presence of an external EM field that is described by the magnetic vector and scalar potentials as $\bm{A}(\bm{r},t)$ and $\phi(\bm{r},t)$. This modified Dirac equation can be expressed by the minimal coupling as \cite{strange98}
\begin{align}
\mathcal{H}_{\rm D}^{\rm EM} & = c\,\bm{\alpha} \cdot \left(\bm{p}-e\bm{A}\right) + \beta m_0c^2 + e\phi\,.
\label{electromag_Dirac}
\end{align}
Note that we have not included magnetic exchange interaction in the following derivation because of its additional complexity. A rigorous calculation of spin dynamics with magnetic exchange for the nonrelativistic spin operator can be found in Refs.\ \cite{Mondal2016,Mondal2018PRB}. 

Now, we perform the FW transformation of the above Hamiltonian and transform the Hamiltonian as an even Hamiltonian \cite{foldy50,strange98,greiner00,bjorken1964relativistic}. The FW transformation can be summarized as $\mathcal{H}_{\rm FW} = e^{iU}\left(\mathcal{H}_{\rm D}^{\rm EM} - i\frac{\partial}{\partial t}\right)e^{-iU} + i\frac{\partial}{\partial t}$, where $U$ defines a unitary operator obtained from the odd terms (i.e., off-diagonal in the particle-antiparticle space) of the Hamiltonian $\mathcal{H}_{\rm D}^{\rm EM}$. The FW transformed Hamiltonian of Eq.\ (\ref{electromag_Dirac}) takes the form \cite{Mondal2017thesis}
\begin{align}
\mathcal{H}_{\rm FW} & = \beta m_0c^2 + \beta\left(\frac{\mathcal{O}^2}{2m_0c^2}-\frac{\mathcal{O}^4}{8m_0^3c^6}\right) + \mathcal{E} -  \frac{1}{8m_0^2c^4}\left[\mathcal{O},\left[\mathcal{O},\mathcal{F}\right]\right]\nonumber\\
& + \frac{\beta}{16m_0^3c^6}\left\{\mathcal{O},\left[\left[\mathcal{O},\mathcal{F}\right],\mathcal{F}\right] \right\}\,,
\end{align}
  with the following definitions of odd and even terms $\mathcal{O} = c\,\bm{\alpha} \cdot \left(\bm{p}-e\bm{A}\right)$ and $\mathcal{E} = e\phi$, respectively. $[A,B]$ defines the commutator, while $\{A,B\}$ defines the anti-commutator for any two given operators $A$ and $B$. Within the FW transformation, the even terms and $i \frac{\partial}{\partial t}$ transform in a similar way, therefore, we introduce a combined term $\mathcal{F} = \mathcal{E} - i \frac{\partial}{\partial t}$ \cite{Silenko2016PRA,Silenko2003,Silenko2016,Silenko2015}. We calculate the corresponding four-component diagonalized Hamiltonian in the particle-antiparticle space that has the form 
\begin{align}
\mathcal{H}_{\rm FW} &  = \beta m_0c^2 +  \frac{\beta\left(\bm{p}-e\bm{A}\right)^2}{2m_0} - \frac{e\beta }{2m_0} \bm{\Sigma}\cdot \bm{B} -  \frac{\beta\left(\bm{p}-e\bm{A}\right)^4}{8m_0^3c^2} + \frac{e\beta }{8m_0^3c^2}\left\{ \left(\bm{p}-e\bm{A}\right)^2,\bm{\Sigma}\cdot \bm{B} \right\}\nonumber\\
& - \frac{\beta e^2B^2}{8m_0^3c^2} - \frac{  e}{8m_0^2c^2}  \bm{\nabla} \cdot \bm{E} + \frac{e}{8m_0^2c^2} \bm{\Sigma} \cdot\left[\left(\bm{p}-e\bm{A}\right) \times \bm{E}
 -  \bm{E} \times \left(\bm{p}-e\bm{A}\right) \right] \nonumber\\
 & - \frac{i e \beta}{16m_0^3c^4}   \bm{\Sigma} \cdot \left[\left(\bm{p}-e\bm{A}\right)\times  \frac{\partial\bm{E}}{\partial t}  +  \frac{\partial\bm{E}}{\partial t}  \times \left(\bm{p}-e\bm{A}\right) \right]\,.
 \label{Full-Hamiltonian}
\end{align}
We have used the following definitions for the Maxwell fields: $\bm{B} = \bm{\nabla}\times \bm{A}$, $\bm{E} = -\frac{\partial \bm{A}}{\partial t} - \bm{\nabla}\phi$.
The above-derived Hamiltonian is very crucial for understanding the light-particle (antiparticle) interaction at low energy excitation. Eq.\ (\ref{Full-Hamiltonian}) can be understood as comprising of nonrelativistic terms and relativistic terms \cite{Mondal2015a}. The first term describes the rest mass energy which has to be subtracted from the total energy in order to obtain the Pauli Hamiltonian for quantum particles. The second term describes the kinetic energy term in the Schr\"odinger equation. The third term is the direct Zeeman coupling of spins with the external magnetic field. The fourth term is the representation of relativistic mass correction terms. The fifth term is an indirect coupling of spins with the external fields. The sixth term is the relativistic correction to the Zeeman coupling. The seventh term explains the Darwin term. The last two terms {represent} the generalized form of spin-orbit coupling. We note that the direct coupling terms  of the spin and the external field are the important ones to describe the corresponding interactions and dynamics \cite{Mondal2018JPCM,hinschberger12}, however, the indirect coupling terms could also be interesting as well \cite{hinschberger12,Zawadzki2005}. We also mention that a full Hamiltonian together with the exchange interaction has also been derived in earlier works where the relativistic corrections to the exchange interactions are obtained \cite{Mondal2015a,Mondal2016,Mondal2018PRB,kraft95,crepieux01}.
The Hamiltonian in Eq.\ (\ref{Full-Hamiltonian}) has been used in calculating the general spin dynamics with the Pauli spin operator \cite{hickey09,Mondal2015a,Mondal2015b,Mondal2016,Mondal2017,Mondal2017Nutation,Mondal2018PRB,Mondal2017thesis,Mondal2018JPCM,Paillard2016SPIE,Mondal2018SPIE,Mondal2019PRB}. We comment that the calculated spin dynamics could explain the precession, spin relaxation of Gilbert type and even nutation dynamics of a single spin \cite{Mondal2017Nutation}. However, the derivation of the spin dynamics has been calculated using a two component extended Pauli Hamiltonian and the nonrelativistic spin operator. Here, our goal is to calculate the spin dynamics from relativistic spin operators.  

The spin-orbit coupling terms can be recast  in a more simplified form by using the well-known Maxwell's equations. Moreover, we can ignore the rest mass energy and constant energy terms in the Hamiltonian of Eq.\ (\ref{Full-Hamiltonian}), because we work out the dynamical equation of motion. The rest of the terms can be simplified to a 
\begin{align}
\mathcal{H}_{\rm FW}^{\rm \prime} &  =   \frac{\beta\left(\bm{p}-e\bm{A}\right)^2}{2m_0} - \frac{e\beta }{2m_0} \bm{\Sigma}\cdot \bm{B} -  \frac{\beta\left(\bm{p}-e\bm{A}\right)^4}{8m_0^3c^2} + \frac{\beta }{8m_0^3c^2}\left\{ \left(\bm{p}-e\bm{A}\right)^2,\bm{\Sigma}\cdot \bm{B} \right\}\nonumber\\
&  - \frac{ e}{8m_0^2c^2}  \bm{\nabla} \cdot \bm{E} - \frac{  \hbar e}{8m_0^2c^2} \bm{\Sigma} \cdot\left[
 2  \bm{E} \times \left(\bm{p}-e\bm{A}\right) -i\hbar \frac{\partial\bm{B}}{\partial t} \right] + \frac{ e \beta}{16m_0^3c^4}   \bm{\Sigma} \cdot  \frac{\partial^2\bm{B}}{\partial t^2}  \,.
\end{align}
Further, we can also ignore the Darwin term in our calculation because the Darwin term involves the density of charges according to the Maxwell theory. As we mentioned earlier, the direct coupling terms provide an opportunity to directly manipulate the spins. 
Therefore, we evaluate in the following the spin dynamics with those terms.

Traditionally one is interested in the direct spin-EM field coupling terms to derive the spin dynamics \cite{Mondal2018JPCM,hinschberger12}. However, the definition of FW or Pryce spin operator suggests that one also needs to consider the terms that do not explicitly depend on the spins. The reason is that the orbital angular momentum enters in the relativistic spin operators in the form of $\bm{p}$. Now, the derivation of spin dynamics follows the time evolution of spin operators that involves the commutators of spin operators with the considered Hamiltonian terms. The commutators of the nonrelativistic Pauli spin operator with spin-independent terms do not contribute to the dynamics. However, for the relativistic spin operator the spin-independent terms have to be considered as well. Therefore, we restrict our derivations to the following FW transformed Hamiltonian 
\begin{align}
\mathcal{H}_{\rm direct}^{\rm spin} &  =  \frac{\beta\left(\bm{p}-e\bm{A}\right)^2}{2m_0}- \frac{e\beta }{2m_0} \bm{\Sigma}\cdot \bm{B} - \frac{   e}{8m_0^2c^2} \bm{\Sigma} \cdot\left[
 2  \bm{E} \times \left(\bm{p}-e\bm{A}\right) -i \frac{\partial\bm{B}}{\partial t} \right]\nonumber\\
 & + \frac{ e \beta}{16m_0^3c^4}   \bm{\Sigma} \cdot  \frac{\partial^2\bm{B}}{\partial t^2}  \,.
 \label{spin-direct-Hamil}
\end{align}
Note that the other relativistic terms will 
contribute to the dynamical equation of motion as well, however, for simplicity of the calculations, we consider only the above-mentioned direct spin-field interaction terms, which are expected to constitute the main contribution. Moreover, Eq.\ (\ref{spin-direct-Hamil}) contains the linear and quadratic interactions in the field, $\bm{A}(\bm{r},t)$. The quadratic terms will become important for the strong field regime \cite{BaukePRA2014,bauke14}. In fact, it has been shown that without these quadratic terms, one cannot describe the spin dynamics qualitatively and quantitatively at the strong field regime \cite{bauke14,Mondal2015b}. In the below, we calculate the spin dynamics with the linear-order interaction terms with the gauge choice, $\bm{A} = \frac{\bm{B}\times \bm{r}}{2}$ which holds for uniform ({\it slowly-varying}) magnetic field {such that $\bm{\nabla}\times \bm{A} = \bm{B}$ and $\bm{\nabla}\cdot \bm{A} = 0$}.    

\section{Derivation of spin dynamics}

\subsection{FW spin operator}
To derive the spin dynamics we calculate the Heisenberg operator dynamics \cite{bjorken1964relativistic}. 

\subsubsection{Free-particle Dirac Hamiltonian.}
The spin dynamics with the free-particle Dirac Hamiltonian is calculated as
\begin{align}
\frac{d\bm{S}_{\rm FW}}{dt} & = \frac{1}{i} \left[\bm{S}_{\rm FW}, \mathcal{H}^{\rm 0}_{\rm D}\right]\nonumber\\
& = \frac{1}{2i}\left[ \bm{\Sigma},\mathcal{H}^{\rm 0}_{\rm D}\right] +\frac{1}{2E_p} \left[\beta\,\bm{p}\times\bm{\alpha}  ,\mathcal{H}^{\rm 0}_{\rm D}\right] - \frac{1}{i} \left[\frac{\bm{p}\times \left( \bm{\Sigma} \times \bm{p}\right)}{2E_p (E_p + m_0c^2)},\mathcal{H}^{\rm 0}_{\rm D}\right]\nonumber\\
& = 0\,.
\label{FW_dynamics1}
\end{align}
 The meaning of Eq.\ (\ref{FW_dynamics1}) is that the FW spin operator is constant of motion {when} a free-particle Dirac Hamiltonian is considered. This result is expected according to the first condition of a proper spin operator \cite{Bauke2014NJP}. Therefore, {the} FW spin operator can be taken as a proper relativistic spin operator.
 
\subsubsection{Dirac Hamiltonian with EM field.}
The spin dynamics for the FW spin operator {for} the Dirac equation {with} an external EM field can be calculated as follows, 
\begin{align}
\frac{d\bm{S}_{\rm FW}}{dt} & = \frac{1}{i} \left[\bm{S}_{\rm FW}, \mathcal{H}^{\rm EM}_{\rm D}\right]\nonumber\\
& = \frac{1}{2i}\left[ \bm{\Sigma},\mathcal{H}^{\rm EM}_{\rm D}\right] +\frac{1}{2E_p} \left[\beta\,\bm{p}\times\bm{\alpha}  ,\mathcal{H}^{\rm EM}_{\rm D}\right] - \frac{1}{i} \left[\frac{\bm{p}\times \left( \bm{\Sigma} \times \bm{p}\right)}{2E_p (E_p + m_0c^2)},\mathcal{H}^{\rm EM}_{\rm D}\right]\nonumber\\
& = - c\bm{\alpha} \times (\bm{p}-e\bm{A}) + \frac{c\beta}{E_p} \bm{p} \times (\bm{p}-e\bm{A}) + \frac{cp^2}{E_p(E_p + m_0c^2)} \bm{\alpha} \times (\bm{p}-e\bm{A})\nonumber\\
& + \frac{ce}{2E_p(E_p + m_0c^2)}\left[(\bm{\alpha}\cdot \bm{r})(\bm{B}\cdot \bm{p})\bm{p} - (\bm{\alpha}\cdot \bm{B})(\bm{r}\cdot \bm{p})\bm{p}\right] \nonumber\\
& + \frac{ce}{4E_p(E_p + m_0c^2)}\Big[\bm{\Sigma}\left[\bm{\alpha}\cdot (\bm{B}\times \bm{p})\right] + (\bm{\Sigma}\cdot \bm{\alpha})(\bm{B}\times \bm{p}) \nonumber\\
&  -\left[\bm{\Sigma}\cdot (\bm{p}\times\bm{\alpha})\right]\bm{B} - (\bm{\Sigma}\cdot \bm{B}) (\bm{p}\times \bm{\alpha})\Big] \,.
\label{FW_dynamics2}
\end{align} 
The derivation followed from the three fundamental commutation relations: $\left[\sigma_i,\sigma_j\right]_- = 2i\epsilon_{ijk} \sigma_k$; $\{ \sigma_i,\sigma_j\}_+ = 2\delta_{ij} I_{2\times 2}$ and $[r_i,p_j] = i \delta_{ij}$, with {$I_{2 \times 2}$ the $2 \times 2$ identity matrix.}
It is evident that when $\bm{A} = \bm{B} = 0$ in Eq.\ (\ref{FW_dynamics2}), the spin dynamics in Eq.\ (\ref{FW_dynamics1}) is recovered. The  meaning of the dynamical terms are explained in the following way. The first term already explains the coupling dynamics for particles and antiparticles. The second term determines the individual dynamics without coupling, however, if $\bm{A} = 0$, this dynamics vanishes because the curl of a gradient is always zero. More importantly, this term does not involve spins because of the fact that the Dirac matrices $\bm{\alpha}$ and $\beta$ anti-commute with each other. The rest of the dynamical terms in Eq.\ (\ref{FW_dynamics2}) are due to the relativistic part of the FW spin operator i.e., the last term of Eq.\ (\ref{FW_spin_operator}). We note that these terms involve, not only, the spins, but also, the product of spins in the dynamics. One of such terms constitutes as $\bm{\Sigma}\cdot \bm{\alpha}$, which can be recast as $\sigma_i^2 = 3\,I_{\rm 2\times 2}$ (assuming Einstein summation convention). 
Therefore, this dynamical term does actually not depend on the spins. Similarly, the other terms containing products of spins can be recast as $\sigma_i\sigma_j = \delta_{ij}I_{\rm 2\times 2} + i\epsilon_{ijk}\sigma_k$, where the first part is again spin-independent, while the second part explicitly depends on spins. 
We conclude for the FW spin-operator dynamics that, along with the spin-dependent dynamics, there are also the spin-independent parts that contribute to the relativistic spin-operator dynamics.

\subsubsection{FW transformation of the Dirac Hamiltonian with EM field.}
Next, we evaluate the FW spin-operator dynamics with the FW transformed Hamiltonian in Eq.\ (\ref{spin-direct-Hamil}). The calculated spin dynamics is
\begin{align}
\frac{d\bm{S}_{\rm FW}}{dt} & = \frac{1}{i} \left[\bm{S}_{\rm FW}, \mathcal{H}_{\rm direct}^{\rm spin}\right]\nonumber\\
& = \frac{1}{2i}\left[ \bm{\Sigma},\mathcal{H}_{\rm direct}^{\rm spin}\right] +\frac{1}{2E_p} \left[\beta\,\bm{p}\times\bm{\alpha}  ,\mathcal{H}_{\rm direct}^{\rm spin}\right] - \frac{1}{i} \left[\frac{\bm{p}\times \left( \bm{\Sigma} \times \bm{p}\right)}{2E_p (E_p + m_0c^2)},\mathcal{H}_{\rm direct}^{\rm spin}\right]\nonumber\\
& = \frac{e\beta }{2m_0} \bm{\Sigma}\times \bm{B} + \frac{1}{E_p}\left[ \frac{\bm{p}\times\bm{\alpha}\left(p^2-e\bm{B}\cdot \bm{L}\right)}{2m_0} + \frac{e (\bm{\Sigma}\cdot \bm{\alpha})  \bm{B}\times \bm{p}}{6m_0} \right] - \frac{e\beta}{4m_0} \frac{\bm{p} \times \left[(\bm{\Sigma} \times\bm{B})\times\bm{p} \right]}{E_p (E_p + m_0c^2)}\nonumber\\
& + \frac{e}{4m_0^2c^2} \left[\bm{\Sigma} \times\left(
   \bm{E} \times \bm{p}\right) + i \frac{ (\bm{p}\times\bm{\alpha}) \times
 (  \bm{E} \times \bm{p})}{E_p} - \frac{\bm{p}\times \left[\left( \bm{\Sigma} \times [ \bm{E} \times \bm{p}] \right)\times \bm{p}\right]}{E_p (E_p + m_0c^2)}\right] \nonumber\\
& -\frac{ie}{8m_0^2c^2}\left[ \bm{\Sigma}\times \dot{\bm{B}}  + i\frac{(\bm{p}\times\bm{\alpha}) \times \dot{\bm{B}}}{E_p}  -  \frac{\bm{\alpha}\times[\dot{\bm{B}}\times (\bm{\Sigma}\times\bm{p})]}{2E_p} +  \frac{\bm{p}\times \left[\left( \bm{\Sigma} \times  \dot{\bm{B}} \right)\times \bm{p}\right]}{E_p (E_p + m_0c^2)}\right] \nonumber\\
& - \frac{ e }{16m_0^3c^4} \left[  \beta\bm{\Sigma} \times  \ddot{\bm{B}}  + \frac{(\bm{\Sigma}\cdot \bm{\alpha}) \ddot{\bm{B}} \times\bm{p}}{3E_p}     +   \beta \frac{\bm{p}\times \left[\left(\bm{\Sigma} \times \ddot{\bm{B}} \right)\times \bm{p} \right] }{E_p (E_p + m_0c^2)}\right]\,.
\label{FW_dynamics3}
\end{align}
As we have started from a semi-relativistic expansion of the Dirac Hamiltonian, it is evidently diagonal in the spin space. However, the calculated spin dynamics suggests that the particle-antiparticle coupling terms (off-diagonal) are nonetheless important, 
when one considers the relativistic FW spin operator.  Furthermore, the spin dynamics shows {the} importance of spin-independent terms in the Hamiltonian in Eq.\ (\ref{spin-direct-Hamil}). The kinetic energy term in Eq.\ (\ref{spin-direct-Hamil}) is explicit spin independent, however, this term contributes to the spin dynamics due to the {form of the} relativistic spin operator. In fact, the commutator $\left[\beta\bm{p}\times\bm{\alpha},\beta \bm{p}\cdot\bm{p}\right]$ leads to an anti-commutator $\left\{\bm{p}\times\bm{\alpha}, \bm{p}\cdot\bm{p}\right\}$ because the Dirac matrices $\bm{\alpha}$ and $\beta$ anti-commute with each other and contribute to the dynamical equation of motion. The diagonal terms in Eq.\ (\ref{FW_dynamics3}) have useful meanings as discussed in the context of magnetization dynamics \cite{Mondal2018PRB,Mondal2017Nutation,Mondal2017thesis}. The first term $\bm{\Sigma}\times \bm{B}$ signifies the precession of a single spin around a field, the terms $\bm{\Sigma} \times\left(
   \bm{E} \times \bm{p}\right)$ and $\bm{\Sigma} \times
   \dot{\bm{B}} $ explains the energy dissipation in terms of damping processes \cite{hickey09,Mondal2016}. The higher-order energy dissipation terms stem from the relativistic parts of the spin operator. These terms can be identified as the last terms in the second and third lines of Eq.\ (\ref{FW_dynamics3}). The other terms in the second and third lines of Eq.\ (\ref{FW_dynamics3}) are evidently off-diagonal, thus, they pertain to the particle-antiparticle interactions. Higher order relativistic spin dynamical terms can be noticed from the last line of Eq.\  (\ref{FW_dynamics3}). Such terms have been associated with  spin dynamics in the inertial regime \cite{Wegrowe2012,Ciornei2011,Wegrowe2016JPCM,Wegrowe2015JAP}, which is a higher-order relativistic spin-orbit coupling effect \cite{Mondal2017Nutation,Mondal2018JPCM}. Note that the dynamical term with $\bm{\Sigma}\cdot \bm{\alpha}$ can be seen as a spin-independent term as described previously. Overall, the spin dynamics with the relativistic FW spin operator exhibits a dynamics that has two contributions: (1) spin-dependent and (2) spin-independent terms.

\subsection{Pryce spin operator}
Another proper relativistic spin operator has been proposed by Pryce \cite{Pryce1948}. 

\subsubsection{Free-particle Dirac Hamiltonian.}
The Pryce spin dynamics with the free-particle Dirac Hamiltonian is calculated as 
\begin{align}
    \frac{d\bm{S}_{\rm Py}}{dt} & = \frac{1}{i} \left[\bm{S}_{\rm Py}, \mathcal{H}^{\rm 0}_{\rm D}\right]\nonumber\\
    & = \frac{1}{2i} \left[\beta \bm{\Sigma}, \mathcal{H}^{\rm 0}_{\rm D}\right] +\frac{1}{2i} \left[\left(1-\beta\right) \frac{(\bm{\Sigma} \cdot \bm{p})\bm{p}}{p^2},\mathcal{H}^{\rm 0}_{\rm D}\right]\nonumber\\
    & = 0\,.
    \label{Pryce_dynamics0}
\end{align}
Again, this result is expected and the Pryce spin operator can be considered as a proper spin operator, similar to the case of the FW spin operator.

\subsubsection{Dirac Hamiltonian with an EM field.}
However, the spin dynamics with the Dirac Hamiltonian in the presence of an EM field is rather different and calculated as
\begin{align}
    \frac{d\bm{S}_{\rm Py}}{dt} & = \frac{1}{i} \left[\bm{S}_{\rm Py}, \mathcal{H}^{\rm EM}_{\rm D}\right]\nonumber\\
    & = \frac{1}{2i} \left[\beta \bm{\Sigma}, \mathcal{H}^{\rm EM}_{\rm D}\right] +\frac{1}{2i} \left[\left(1-\beta\right) \frac{(\bm{\Sigma} \cdot \bm{p})\bm{p}}{p^2},\mathcal{H}^{\rm EM}_{\rm D}\right]\nonumber\\
    & =  
    \frac{ec}{4p^2} (\bm{\Sigma}\times \bm{B}) \bm{\alpha}\cdot \bm{p} + \frac{ec}{2p^2}\left[(\bm{\alpha}\cdot \bm{r})(\bm{B}\cdot \bm{p})\bm{p} - (\bm{r}\cdot \bm{p})(\bm{\alpha}\cdot \bm{B})\bm{p} \right]\,.
    \label{Pryce_dynamics1}
\end{align}
In the above derivation, the first commutator $[\beta \bm{\Sigma}, c\bm{\alpha}\cdot\bm{p}]$ exactly cancels the last commutator $[\beta \frac{(\bm{\Sigma}\cdot \bm{p})\bm{p}}{p^2}, c\bm{\alpha}\cdot\bm{p}]$. Therefore, only the remaining commutator $[ \frac{(\bm{\Sigma}\cdot \bm{p})\bm{p}}{p^2}, c\bm{\alpha}\cdot\bm{p}]$ contributes to the spin dynamics.  
Note that in the absence of the EM field i.e., $\bm{B} = 0$, the dynamics in Eq.\ (\ref{Pryce_dynamics1}) recovers the spin dynamics for a free Dirac particle in Eq.\ (\ref{Pryce_dynamics0}). It is interesting to point out that the dynamics in Eq.\ (\ref{Pryce_dynamics1}) contains only the off-diagonal elements in the matrix formalism. The latter means that this dynamics is governed by the coupling between the particles and antiparticles which comes from the Dirac Hamiltonian itself, the term $\bm{\alpha}\cdot \bm{p}$. This feature of the Pryce spin dynamics {stands} in contrast {to} the FW spin dynamics in Eq.\ (\ref{FW_dynamics2}), where {both} diagonal and off-diagonal terms contribute. In fact, the FW spin dynamics contain terms with only diagonal {contributions}. The first term in Eq.\ (\ref{Pryce_dynamics1}) is notably off-diagonal and can be represented by $\sigma_i\sigma_j$. Following the similar argument, this term can be split into a spin-independent part and a spin-dependent part. Thus, the Pryce spin dynamics contains also spin dependent and independent contributions, like
the FW spin dynamics as discussed earlier. .

\subsubsection{FW transformation of the Dirac Hamiltonian with an EM field.}
Next, we calculate the spin dynamics from the  transformed Hamiltonian in Eq.\ (\ref{spin-direct-Hamil}). The derived dynamical equation is 
\begin{align}
    \frac{d\bm{S}_{\rm Py}}{dt} & = \frac{1}{i} \left[\bm{S}_{\rm Py}, \mathcal{H}_{\rm direct}^{\rm spin}\right]\nonumber\\
    & = \frac{1}{2i} \left[\beta \bm{\Sigma}, \mathcal{H}_{\rm direct}^{\rm spin}\right] +\frac{1}{2i} \left[\left(1-\beta\right) \frac{(\bm{\Sigma} \cdot \bm{p})\bm{p}}{p^2},\mathcal{H}_{\rm direct}^{\rm spin}\right]\nonumber\\
& = \frac{e}{2m_0} \bm{\Sigma}\times\bm{B} + \frac{e\beta(1-\beta)}{4m_0p^2}  \bm{\Sigma}\times\left[\bm{p}\times(\bm{B}\times\bm{p})\right]  + \frac{e\beta}{4m_0^2c^2}  \bm{\Sigma} \times\left(
  \bm{E} \times \bm{p} \right)\nonumber\\
 &  + \frac{e(1-\beta)}{8m_0^2c^2p^2}\left[ (\bm{\Sigma}\cdot\bm{p}) (\bm{\Sigma}\cdot \dot{\bm{B}})\bm{p} - (\bm{\Sigma}\cdot \bm{p})(\bm{L}\cdot\dot{\bm{B}})\bm{p} - \frac{\Sigma^2\bm{p}+(\bm{\Sigma}\cdot \bm{p})\bm{\Sigma}}{2}(\dot{\bm{B}}\cdot\bm{p})\right]\nonumber\\
& -\frac{ie\beta}{8m_0^2c^2} \left[  \bm{\Sigma} \times  \dot{\bm{B}} +\beta(1-\beta)\frac{[(\bm{\Sigma}\times\dot{\bm{B}})\cdot\bm{p}]\bm{p}}{p^2} \right]\nonumber\\
& - \frac{e}{16m_0^3c^4}   \left[\bm{\Sigma} \times  \ddot{\bm{B}} +\beta(1-\beta)  \frac{[(\bm{\Sigma}\times\ddot{\bm{B}})\cdot\bm{p}]\bm{p}}{p^2}\right]\,.
\label{Pryce_dynamics2}
\end{align}
As we have started from a diagonalized Hamiltonian and {the} Pryce spin operator which is diagonal, too, all the derived dynamical terms are diagonal as well. This means that the corresponding dynamics only describes the particles and antiparticles, not the coupling between them. To derive such dynamics, one has to note that the kinetic energy does commute with the first term of the Pryce spin operator in Eq.\ (\ref{Pryce_spin_operator}), however, it does not commute with the second term because the latter contains the momentum operator as well. Such non-commutator implies that {\it not only} the spin, {\it but also} the orbital momentum  contributes to the relativistic spin dynamics through the spin-orbit coupling-like mechanisms that is considered in the relativistic spin operator of Pryce type. 
The dynamical terms in Eq.\ (\ref{Pryce_dynamics2}) can be related to the similar terms as was derived in Eq.\ (\ref{FW_dynamics3}). For example, the first term in Eq.\ (\ref{Pryce_dynamics2}) describes the spin precession around a field, the third term and the first terms of third  line in Eq.\ (\ref{Pryce_dynamics2}) explain the energy dissipation from spin to other degrees of freedom. The first term of the last line in Eq.\ (\ref{Pryce_dynamics2}) accounts for the spin dynamics {in} the inertial regime. The other remaining terms in Eq.\ (\ref{Pryce_dynamics2}) do not directly correspond to the FW dynamics in Eq.\ (\ref{FW_dynamics3}). However, they derive from the relativistic part of the Pryce spin operator. Thus, they contain either $(1-\beta)$ or $\beta(1-\beta)$ as appear in Eq.\ (\ref{Pryce_dynamics2}).            

\section{Summary and Discussions}
Traditionally, the spin dynamics is derived for the nonrelativistic spin operator (see, e.g., \cite{Mondal2018PRB}). Here, we have derived the spin dynamics with relativistic spin operators. We have used three different Hamiltonians to derive {the corresponding} spin dynamics: (1) free-particle Dirac Hamiltonian, (2) Dirac Hamiltonian in an EM environment, (3) diagonalized Dirac Hamiltonian in the presence of an EM field. The relativistic spin dynamics is a constant of motion when the free-particle Dirac Hamiltonian is considered. This result however only holds for relativistic spin operators of FW and Pryce type, making them ideal candidates for proper relativistic spin operators. These two relativistic spin operators are however very different:
the FW spin operator has diagonal and off-diagonal elements in spin space, whereas,
the Pryce operator has only diagonal elements. These spin operators involve not only spin angular momentum in terms of Pauli spin matrices, but also, the orbital angular momentum in terms of momentum operator $\bm{p}$. The derived dynamics of these operators in an EM field provides two important informations: (1) the particle-antiparticle coupling terms contribute to the spin dynamics, even if one starts with a diagonalised Hamiltonian, (2) there exist two separate parts (spin-dependent and spin-independent terms) of the derived spin dynamics. We note that some dynamical terms appear in both the FW and Pryce spin dynamics in similar way, however, due of the relativistic spin operators' construction, additional terms exist. The derived dynamics reveals that coupling of the orbital angular momentum with spin contributes to the spin dynamics, moreover, a few dynamical terms {\it only} depend on the orbital angular momentum.         

\begin{figure}[!h]
    \centering
  \includegraphics[scale = 1.0]{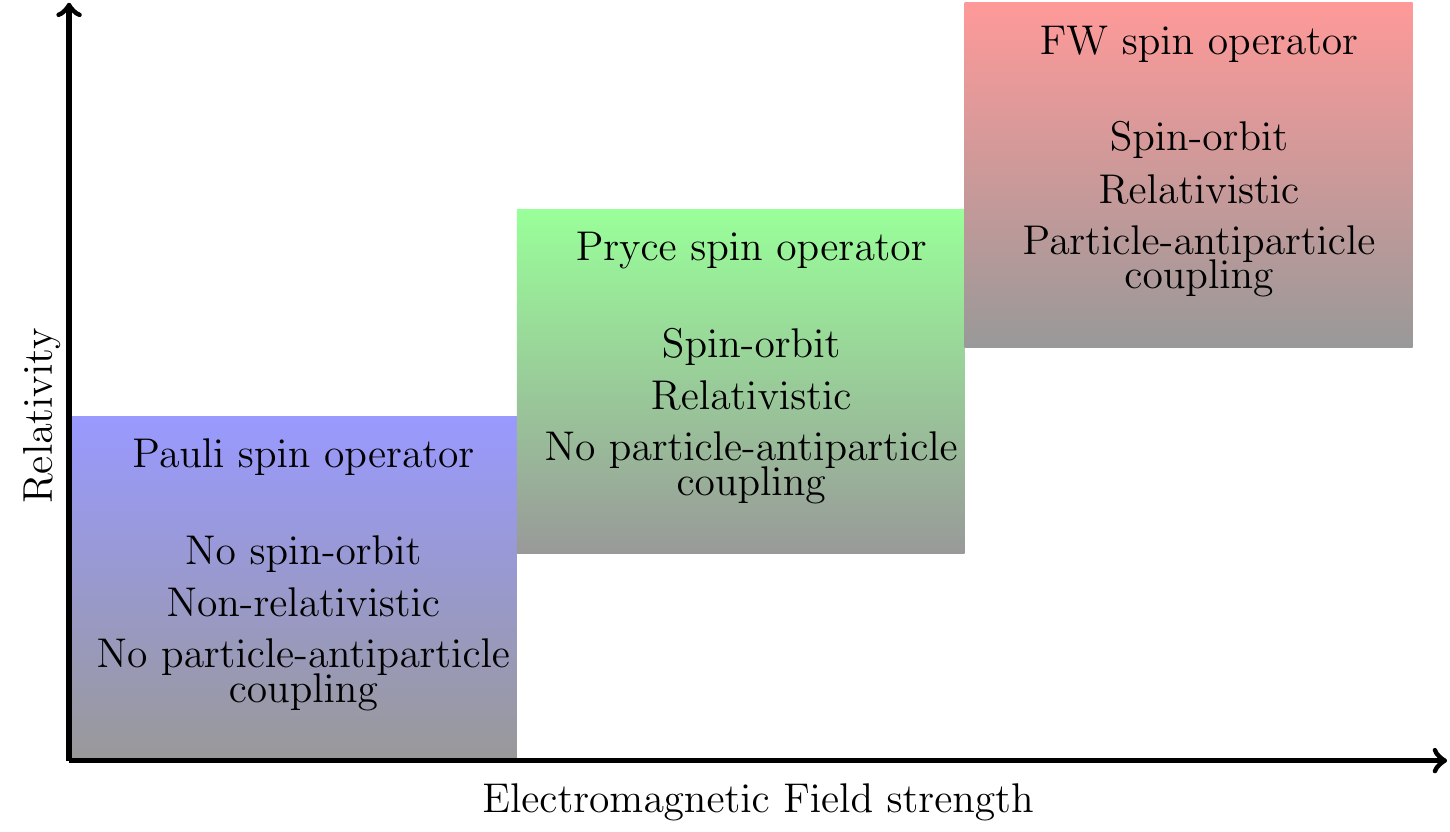}
    \caption{(Color Online): A schematic for operational spin dynamics of three different spin operators at varying EM
        field strengths. For the weak field strength and nonrelativistic regime, the Pauli spin operator is enough to describe the corresponding spin dynamics \cite{Mondal2016}, however, in the relativistic regime and for intermediate to strong field strengths the Pryce and FW spin operators are respectively suitable.
    }
    \label{comparison}
\end{figure}

Several terms in the derived spin dynamics equations
of the two considered proper spin operators are rather
distinct. The FW spin operator has diagonal and off-diagonal components which means it accounts for the coupling terms in the particle-antiparticle Hilbert space. 
When we compare the two equations for spin motion,
Eq.\ (\ref{FW_dynamics3}) for FW dynamics and Eq.\ (\ref{Pryce_dynamics2}) for Pryce dynamics, which have been derived from the same Hamiltonian, we observe that 
the Pryce dynamics in Eq.\ (\ref{Pryce_dynamics2}) is diagonal and does not involve such coupling terms. In fact, the Pryce dynamics involves terms which have ($1-\beta$) that translates to zero contribution for the upper component in $2\times 2$ formalism. Therefore, the $2\times 2$ Pryce dynamics recovers exactly the same dynamical terms as the Pauli spin dynamics \cite{Mondal2017Nutation}. As already mentioned, the FW dynamics in Eq.\ (\ref{FW_dynamics3}) contains diagonal as well as off-diagonal terms. To achieve a $2\times 2$  electron spin dynamics, one has to diagonalize.
Even then, the additional terms appear apart from the standard Pauli spin dynamics. Moreover, the additional terms account for spin angular momentum and orbital contributions as well. Therefore, one can conclude that for the spin dynamics in an applied EM field, the two spin operators have their own validity regime. We thus consider
three operational field regimes: weak, intermediate, and strong. In the weak field regime, the Pauli spin operator can describe the spin dynamics, while, for an intermediate field regime where the spin-orbit coupling is important, the Pryce spin operator seems to describe the proper spin dynamics. However, in the stronger field regime, where the spin-orbit and relativistic particle-antiparticle couplings are present, the FW spin operator suits the best for describing the spin dynamics. The derived operational spin dynamics regimes of the Pauli, Pryce and FW spin operators are schematically summarized in Fig.\ \ref{comparison}.

\section{Acknowledgments}
R.M.\ acknowledges Arnab Rudra for fruitful discussions and the Alexander von Humboldt foundation for the postdoctoral fellowship and Zukunftskolleg at Universit\"at Konstanz (grant No.\ P82963319) for financial support.  P.M.O.\ acknowledges support from the Swedish Research Council (VR) and the K.\ and A.\ Wallenberg Foundation (grant No.\ 2015.0060).

\bibliographystyle{iopart-num}

\providecommand{\newblock}{}

\end{document}